\documentclass[cits]{PoS}

\title{Nucleosynthesis in dynamical and torus ejecta of compact binary mergers}

\ShortTitle{Nucleosynthesis in compact binary mergers}

\author{\speaker{Oliver Just},$^{1,2}$ \\
        E-mail: \email{ojust@mpa-garching.mpg.de}}

\author{
  A.~Bauswein,$^3$  R.~Ardevol Pulpillo,$^{1,4}$ S.~Goriely$^5$ and H.-T.~Janka$^1$ \\
   \llap{$^1$} Max-Planck-Institut f\"ur Astrophysik, Postfach 1317, 85741 Garching, Germany \\
   \llap{$^2$} Max-Planck/Princeton Center for Plasma Physics (MPPC) \\
   \llap{$^3$} Department of Physics, Aristotle University of Thessaloniki, 54124 Thessaloniki, Greece \\
   \llap{$^4$} Physik Department, Technische Universit\"at M\"unchen, James-Franck-Stra\ss e 1, 85748 Garching, Germany \\
   \llap{$^5$} Institut d'Astronomie et d'Astrophysique, CP-226, Universit\'e Libre de Bruxelles, 
    1050 Brussels, Belgium
  }

\newcommand{\aap}{A\&A}
\newcommand{\apjl}{ApJL}

\newcommand{\apj}{ApJ}
\newcommand{\aanda}{A\&A}
\newcommand{\prd}{Phys. Rev. D}
\newcommand{\prc}{Phys. Rev. C}

\newcommand{\mnras}{MNRAS}

\newcommand{\pasa}{PASA}

\def\ga{\,\,\raise0.14em\hbox{$>$}\kern-0.76em\lower0.28em\hbox
{$\sim$}\,\,}
\def\la{\,\,\raise0.14em\hbox{$<$}\kern-0.76em\lower0.28em\hbox
{$\sim$}\,\,}

\abstract{We present a comprehensive study of r-process element nucleosynthesis in the ejecta of
  compact binary mergers (CBMs) and their relic black-hole (BH)-torus systems. The evolution of the
  BH-accretion tori is simulated for seconds with a Newtonian hydrodynamics code including
  viscosity, pseudo-Newtonian gravity for rotating BHs, and an energy-dependent two-moment closure
  neutrino transport scheme. The investigated cases are guided by relativistic double neutron star
  (NS-NS) and NS-BH merger models. Our nucleosynthesis analysis includes the dynamical ejecta
  expelled during the CBM phase and the neutrino and viscously driven outflows of the relic BH-torus
  systems. While typically $\sim 20-25\,\%$ of the initial torus mass are lost by viscously driven
  outflows, neutrino-powered winds contribute at most another $\sim 1\,\%$. Since BH-torus ejecta
  possess a wide distribution of electron fractions and entropies, they produce heavy elements from
  A~80 up to the actinides, with relative contributions of $A>130$ nuclei being subdominant. The
  combined ejecta of CBM and BH-torus phases can reproduce the solar abundances amazingly well for
  $A>90$. Varying contributions of the torus ejecta might account for observed variations of lighter
  elements with $40<Z<56$ relative to heavier ones, and a considerable reduction of the prompt
  ejecta compared to the torus ejecta, e.g. in highly asymmetric NS-BH mergers, might explain the
  composition of heavy-element deficient stars.}

\FullConference{XIII Nuclei in the Cosmos\\
		 7-11 July, 2014\\
		 Debrecen, Hungary}

\begin{document}

\begin{figure}
  \hspace{-5mm}
  \includegraphics[scale=0.38,angle=0]{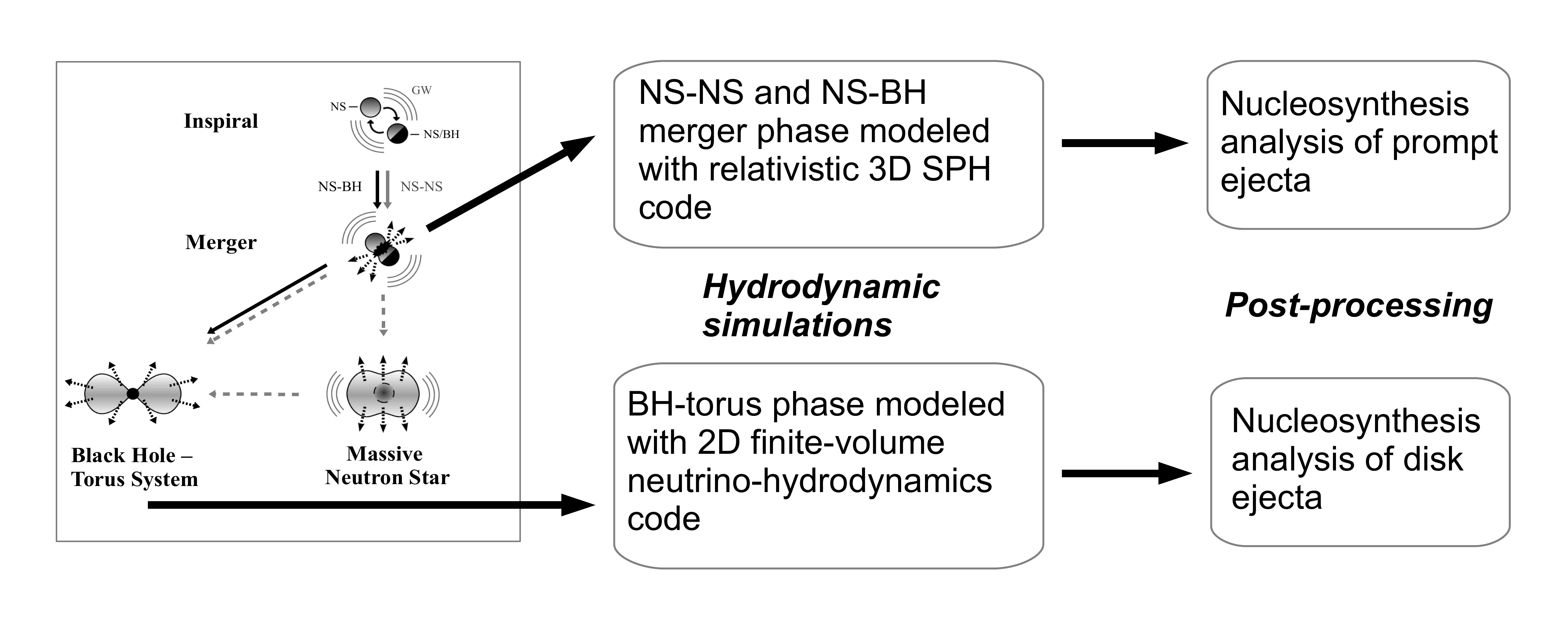}
  \vspace{-1.2cm}
  \caption{Schematical overview of our study. See main text for details.}
  \vspace{-0.3cm}
  \label{fig:overview}
\end{figure}

\section{Motivation and Overview}

The astrophysical site(s) for the production of r-process elements is (are) still mysterious. Recent
studies suggest that instead of or in addition to the long favored core-collapse supernovae the
outflows from compact binary (i.e. double neutron-star, NS-NS, or neutron-star black-hole, NS-BH)
mergers could provide suitable conditions to allow for the r-process. Most existing nucleosynthesis
studies \cite{Freiburghaus1999, Goriely2011, Roberts2011, Korobkin2012, Bauswein2013, Wanajo2014a}
only consider the prompt (dynamical) ejecta launched during the merger phase in a NS-NS
configuration. However, also the dynamical ejecta from NS-BH mergers as well as the ejecta stemming
from the remnants of both merger types need to be taken into account in order to understand the full
chemogalactic fingerprint of compact object mergers. In both merger types, the merger remnant can
consist of a BH surrounded by a more or less massive torus of neutron-star debris. In
\cite{Just2014b} we conducted a comprehensive analysis of the nucleosynthetic output associated with
the ejecta from the merging phase of NS-NS and NS-BH binaries {\em and} with the outflows of the
subsequent long-time evolution of relic BH-torus systems. We summarize the individual steps of this
study in Fig.~\ref{fig:overview}. The compact binary mergers are simulated with a relativistic
smooth-particle-hydrodynamics code \cite{Oechslin2007}, while the BH-torus modeling is performed
with a Eulerian finite-volume Godunov-type scheme \cite{Obergaulinger2008}, supplemented by a
shear-viscosity treatment with a Shakura-Sunyaev $\alpha$-prescription for the dynamic viscosity
\cite{Shakura1973}. For the long-time evolution we employ a pseudo-Newtonian approximation of the
gravity potential for the rotating relic BHs \cite{Artemova1996}. In the time-dependent BH-torus
modeling we apply, for the first time, detailed energy-dependent and velocity-dependent 2D neutrino
transport based on a new two-moment closure scheme \cite{Just2014}, which allows us to determine the
neutrino-driven wind and the neutron-to-proton ratio in the disk outflows with higher accuracy than
in previous simulations. All simulations include microphysical treatments of the gas equation of
state (EOS).

Our nucleosynthesis calculations are carried out in a post-processing step of the ejecta produced by
the hydrodynamical models, using a full r-process network including all relevant nuclear reactions
\cite{Goriely2008,Goriely2010,Xu2013}. For the combined analysis we pick cases from larger sets of
NS-NS merger models, which lead to prompt or slightly delayed BH formation, and NS-BH merger models
on the one side and BH-torus models on the other side such that the macroscopic system parameters
(BH and torus masses and BH spins of the merger remnants) match roughly on both sides.

\begin{figure}
  \includegraphics[width=70mm]{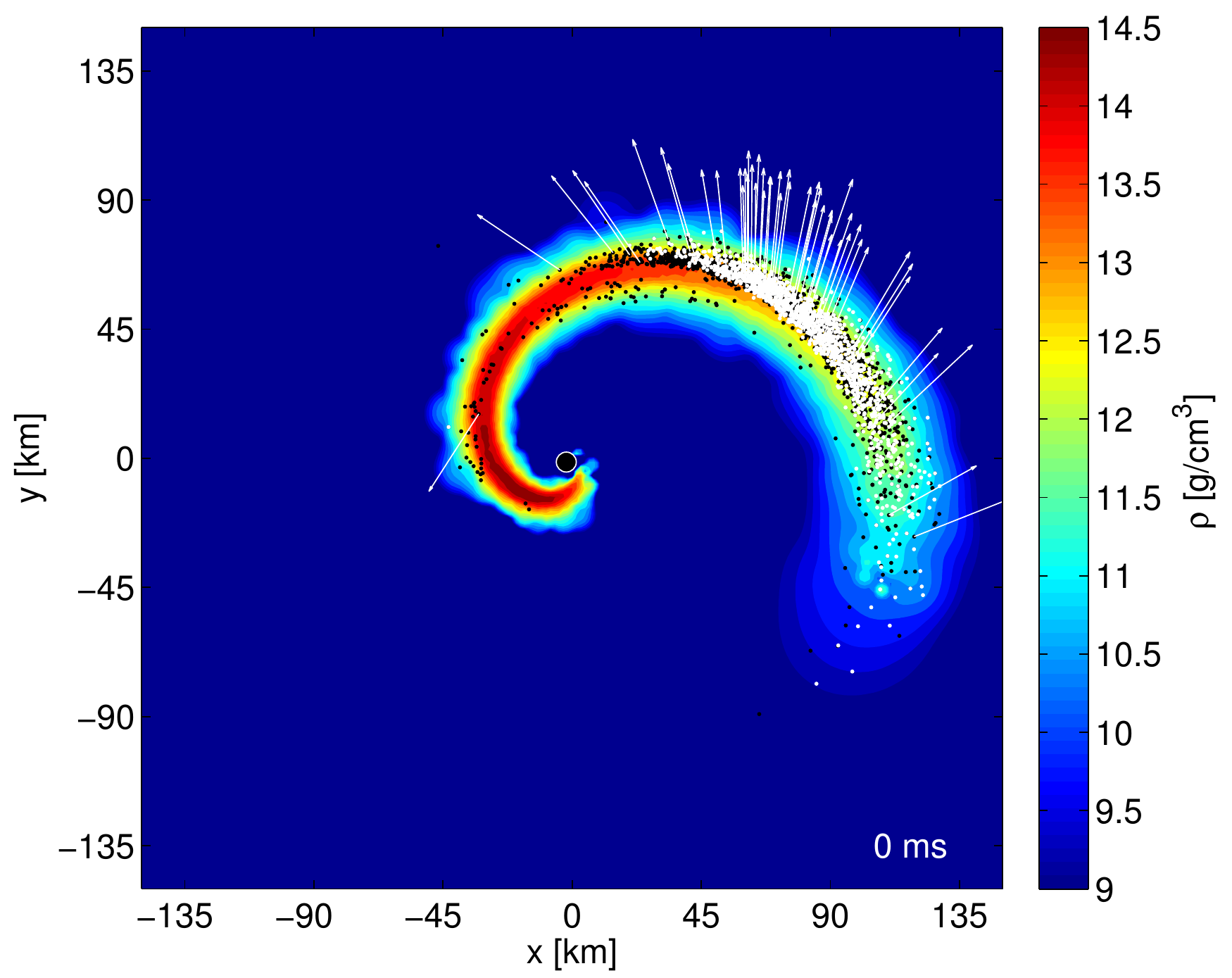}
  \includegraphics[width=80mm]{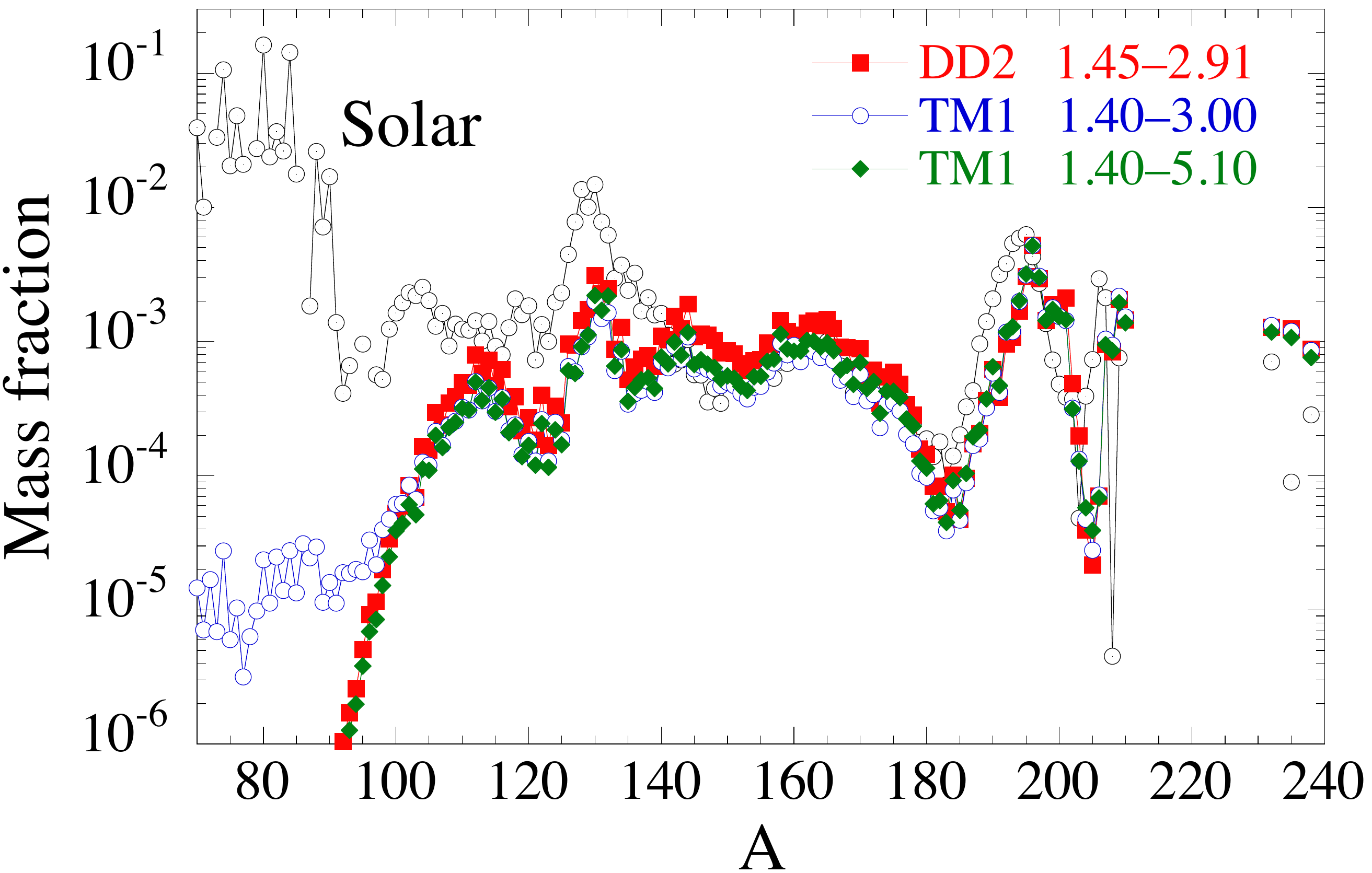}
  \caption{\emph{Left:} Rest-mass density in the equatorial plane during the merger of a
    1.4\,$M_\odot$ NS with a 5.1\,$M_\odot$ BH. Dots denote particles (projected into the equatorial
    plane) which are or will become gravitationally unbound while arrows indicate the corresponding
    velocities. The visualization tool SPLASH was used to convert SPH data to grid data
    \cite{Price2007}. \emph{Right:} Abundance distributions as functions of the atomic mass for the
    dynamical ejecta of three NS-BH merger cases. Each binary system is characterized in the legend
    by the EOS used in the simulation and the mass (in $M_\odot$) of the NS and BH,
    respectively. All distributions are normalized to the same A = 196 abundance. The dotted circles
    show the solar r-abundance distribution \cite{Goriely1999a}.}
  \label{fig:merger_ejecta}
\end{figure}

\section{Ejecta from the Merger Phase}

Consistent with previously published relativistic results \cite{Bauswein2013, Foucart2014,
  Wanajo2014a}, the ``dynamical'' ejecta of NS-NS and NS-BH mergers, which are expelled within
milliseconds of the collision of the two binary components, were found to possess similar average
properties, namely expansion velocities of 0.2--0.4\,$c$, electron fractions below $\sim 0.1$, and
entropies per baryon of a few $k_\mathrm{B}$. The considered NS-NS mergers produce
$\sim$0.004--0.021\,$M_\odot$ of ejecta, whereas the NS-BH mergers eject significantly larger
masses, 0.035--0.08\,$M_\odot$, with very low entropies ($\la 1\,k_\mathrm{B}$ per nucleon), because
this matter is not shock heated as in NS-NS collisions, but originates mostly from the outer tail of
the tidally stretched NS at its final approach to the BH. Mass lost in NS-BH mergers is also
expelled much more asymmetrically than in the case of NS-NS mergers (see left plot in
Fig.~\ref{fig:merger_ejecta}): Corresponding hemispheric asymmetry parameters (mass difference
between dominant ejecta hemisphere and opposite hemisphere, divided by total ejecta mass) are a few
per cent for symmetric NS-NS mergers and 15--30\% for strongly asymmetric ones, but 0.93--0.98 for
NS-BH mergers.

Since the high neutron excess, thermodynamic properties, and expansion timescale are very similar,
the ejecta of NS-NS mergers as well as those of NS-BH mergers are sites of robust production of
r-nuclei with $A\ga 140$ and abundances close to the solar distribution (see right plot in
Fig.~\ref{fig:merger_ejecta} for typical abundance pattern resulting in the NS-BH case). This result
holds basically independently of the considered nuclear EOS and the exact binary parameters and
confirms the findings of previous studies based on relativistic merger simulations
\cite{Bauswein2013, Hotokezaka2013, Wanajo2014a}.

\begin{figure}
  \includegraphics[width=60mm]{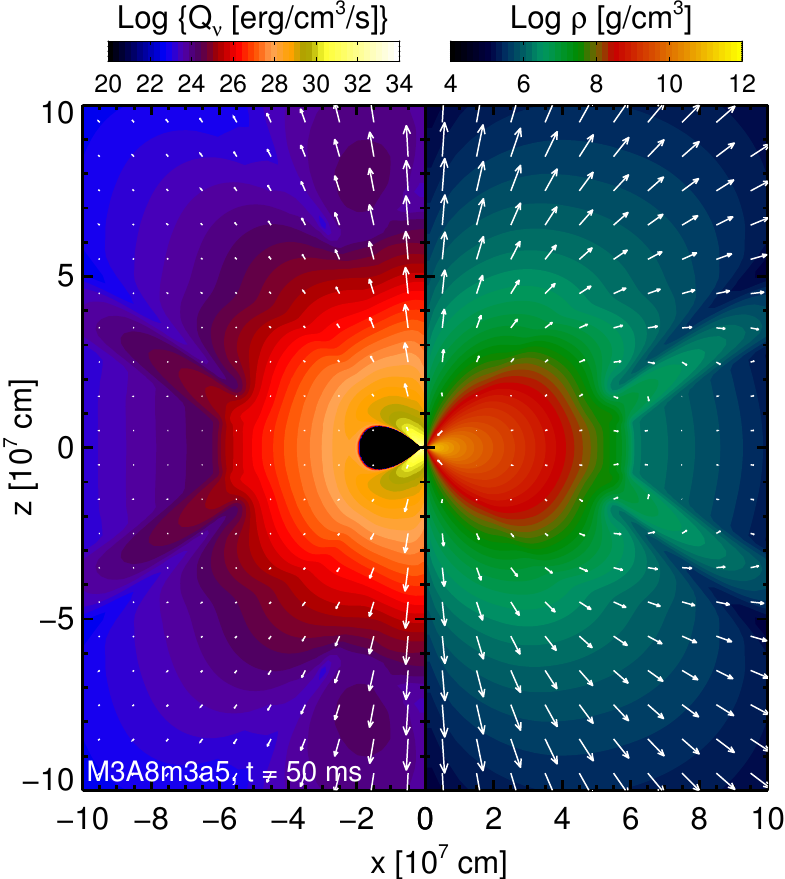}
  \hspace{3mm}
  \includegraphics[width=85mm]{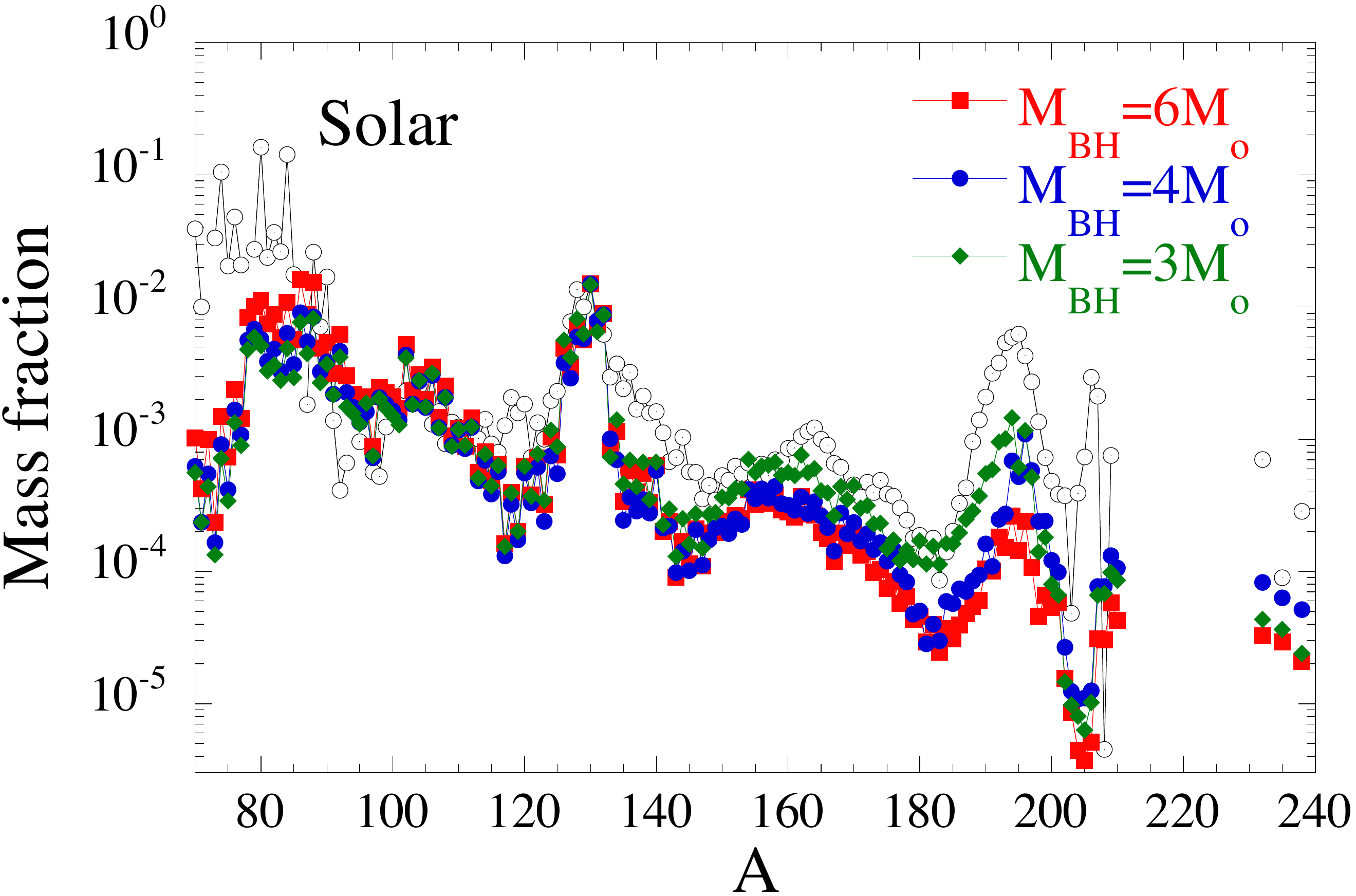}
  \caption{\emph{Left:} Snapshot of a BH-torus model with a 3\,$M_\odot$ central BH with spin
    parameter 0.8 and an initial torus mass of 0.3\,$M_\odot$ at about 50\,ms after the merger. On
    the left side the net neutrino heating rate due to $\beta$-processes and neutrino pair
    annihilation is color coded and overlaid with arrows indicating the energy-integrated flux
    density of electron neutrinos multiplied with the squared radius. On the right side the density
    is color coded and overlaid with velocity arrows. \emph{Right:} Abundance distributions as
    functions of the atomic mass for three BH-torus systems with BH masses of $M_{\mathrm{BH}}= 3$,
    4 and 6\,$M_\odot$ and the same 0.3\,$M_\odot$ tori. All distributions are normalized to the
    same solar $A=130$ abundance. The dotted circles show the solar r-abundance distribution
    \cite{Goriely1999a}.}
  \label{fig:torus_ejecta}
\end{figure}

\section{Ejecta from the BH-Torus Phase}

The relic BH-torus systems lose mass in neutrino-driven baryonic winds \cite{Wanajo2012} and in
outflows triggered by viscous energy dissipation and angular momentum transport \cite{Fernandez2013,
  Metzger2014, Fernandez2014}. We find that the viscously driven ejecta amount up to, fairly model
independent, 19--26\% of the initial torus mass (for our BH spins of $A_\mathrm{BH} = 0.8$) and
dominate the neutrino-driven ejecta by far. The neutrino-driven wind depends extremely sensitively
on the BH mass and on the torus mass, which determines the neutrino luminosities. The neutrino
energy loss rates can reach several 10$^{52}$\,erg\,s$^{-1}$ up to more than
10$^{53}$\,erg\,s$^{-1}$ for each of $\nu_e$ and $\bar\nu_e$ for a few 100\,ms. The neutrino-driven
ejecta carry away up to about one per cent of the initial torus mass, but their mass can also be
orders of magnitude lower. The maximum masses that can be associated with ejecta driven by neutrino
heating are 2.5--3.5$\times 10^{-3}$\,$M_\odot$ in the case of high neutrino luminosities for
typical durations of fractions of a second up to $\sim$1\,s. This number agrees with the lower limit
of the expelled mass computed by \cite{Perego2014a} for a hypermassive NS as merger remnant, and the
corresponding mass-loss rate is roughly compatible with the values obtained for neutrino winds of
proto-neutron stars in supernova cores \cite{Qian1996}. Neutrino heating, however, inflates the
outer layers of the torus and has a positive feedback on the viscously driven mass ejection on the
level of several per cent of the torus mass (or up to $\sim$20\% of the viscous-outflow mass).

Neutrino-driven torus winds exhibit characteristic properties which distinguish them from the
viscously triggered outflow, namely the tendency of higher mean entropies, higher electron
fractions, and larger expansion velocities. Moreover, the neutrino wind is strongest at early times
and at intermediate latitudes (around 45$^\circ$ away from the equatorial plane), and the entropy,
electron fraction, and velocity exhibit a strong pole-to-equator variation with higher values
towards the poles. In contrast, viscously driven ejecta develop on longer timescales, are more
spherical, and their properties vary little with angular direction.

Because of their greater neutron excess, viscously-driven ejecta allow for a much stronger r-process
than the neutrino wind. The combination of both components is far dominated by the viscous
contribution and matches the solar abundance pattern for all nuclear mass numbers $A \ga 90$ fairly
well. The abundance pattern for $A \le 132$ is comparatively uniform, but the strength of the third
abundance peak decreases with higher BH mass (see Fig.~\ref{fig:torus_ejecta}), and the relative
yields of low-mass ($A \la 130$) and high-mass ($A \ga 130$) components depend on the value of the
dynamic viscosity and the detailed treatment of the viscosity terms in the hydrodynamics equations.

\begin{figure}
  \includegraphics[width=75mm]{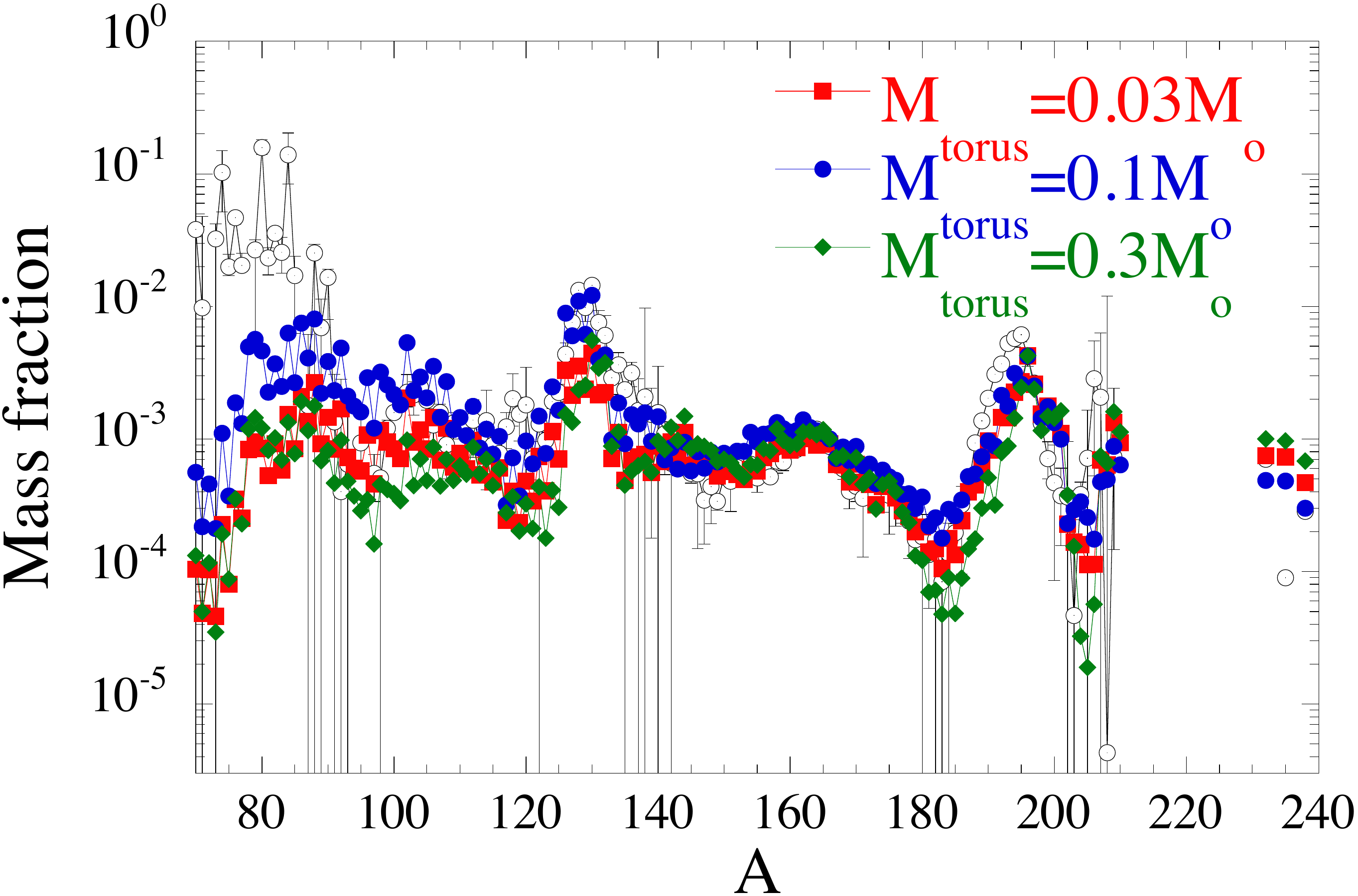}
  \hspace{3mm}
  \includegraphics[width=70mm]{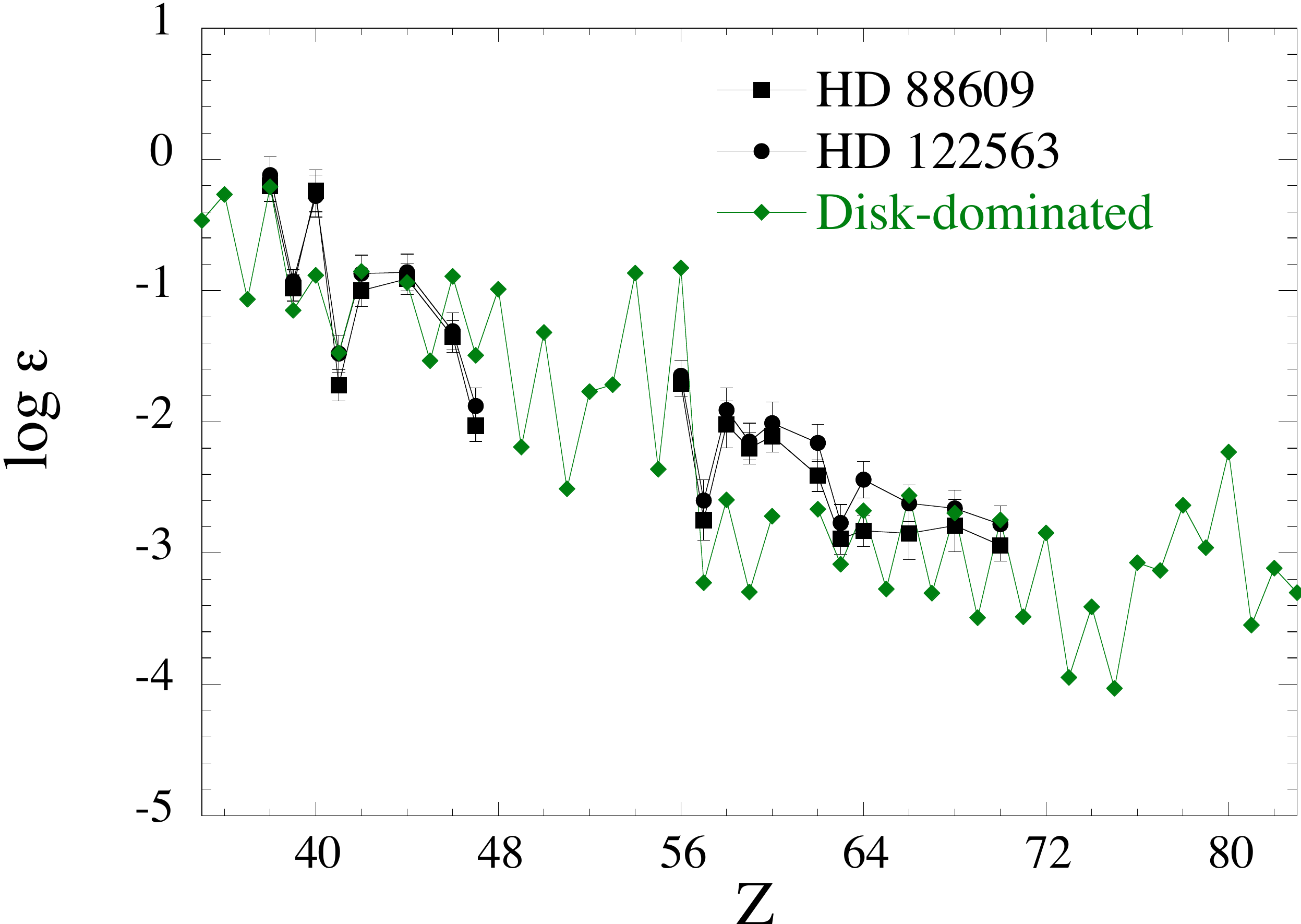}
  \caption{\emph{Left:} Abundance distributions as functions of the atomic mass for three combined
    systems (merger model plus remnant model) corresponding to models with the indicated torus
    masses. All distributions are normalized to the same solar $A=196$ abundance. The dotted circles
    show the solar r-abundance distribution \cite{Goriely1999a}. \emph{Right:} Comparison between
    the HD88609 and HD122563 elemental abundances (in log\,$\varepsilon$ scale) and those estimated
    for the combination of a merger model and a BH-torus model, when the ejected mass of the merger
    model is assumed to be 100 times smaller than the one coming from the BH-torus model. The
    calculated distribution is normalized to the Sr abundance of HD88609.}
  \label{fig:combined_ejecta}
\end{figure}

\section{Combined Ejecta and Observational Implications}

The consistently mass-weighted combination of the prompt ejecta from the merger models and the
secular ejecta from the BH-torus models can reproduce amazingly well the solar r-abundance pattern
in the range $90\la A < 240$ and therefore also that seen in ultra-metal-poor stars (cf. left plot
in Fig.~\ref{fig:combined_ejecta}). In particular, the BH-torus outflows are able to well fill the
region $A \la 140$, where the prompt merger ejecta underproduce the nuclei (but see
\cite{Wanajo2014a}). Since the relative yields of the relic BH-torus systems for $A \la 130$ and
$A\ga 130$ nuclei depend sensitively on the system parameters, whereas the $A\ga 140$ species are
created with a robust pattern during the binary merging phase, we expect a larger variability in the
low-$A$ regime than for high mass numbers.

Another interesting possibility is connected to the fact that the mass ejection during NS-BH mergers
shows extreme spatial asymmetry (corresponding to asymmetry parameters $\ga$0.95) but the mass loss
from their BH-torus remnants is much more isotropic. This can lead to a strong suppression of the
dynamical ejecta component relative to the torus outflow in observer directions pointing away from
the hemisphere that receives most of the expelled matter of the disrupted NS. Essentially pure
BH-torus ejecta of some of our models can reasonably well match the abundance distributions observed
in heavy-element deficient metal-poor stars like HD88609 and HD122563 (right plot in
Fig~\ref{fig:combined_ejecta}).

Since the major part of the torus ejecta ends up in forming $A\la 130$ material, the additional mass
loss of the merger remnants does not alter event-rate estimates based on comparing yields of heavy
r-nuclei in the ejecta of the binary-merger phase with the r-process abundances in our Galaxy
\cite{Goriely2011,Bauswein2014}.

\section*{Acknowledgements}
At Garching, this research was supported by the Max-Planck/Princeton Center for Plasma Physics
(MPPC) and by the Deutsche Forschungsgemeinschaft through the SFB-TR7 and the Cluster of Excellence
EXC 153. AB is a Marie Curie Intra-European Fellow within the 7th European Community Framework
Programme (IEF 331873). SG acknowledges financial support from FNRS (Belgium). We are also grateful
for computational support by the Center for Computational Astrophysics (C2PAP) at the Leibniz
Rechenzentrum (LRZ) and by the Rechenzentrum Garching (RZG).


\providecommand{\href}[2]{#2}\begingroup\raggedright\endgroup

\end{document}